\documentclass[runningheads]{llncs}
\usepackage[T1]{fontenc}
\usepackage{graphicx}
\usepackage{amsmath} 
\usepackage{amsfonts} 
\usepackage{graphicx}  
\usepackage{fontawesome} 
\usepackage{subcaption}  
\usepackage[hidelinks]{hyperref}

\let\svthefootnote\thefootnote
\newcommand\freefootnote[1]{%
  \let\thefootnote\relax%
  \footnotetext{#1}%
  \let\thefootnote\svthefootnote%
}

\newcommand{\PhenDiff}{\textbf{PhenDiff}}

\begin{document}
\title{PhenDiff: Revealing Subtle Phenotypes with Diffusion Models in Real Images}
\titlerunning{PhenDiff}

\author{Anis Bourou\inst{1,2*} \and
Thomas Boyer\inst{1*} \and
Marzieh Gheisari\inst{1} \and
Kévin Daupin\inst{2}  \and
Véronique Dubreuil\inst{2}  \and
Aurélie De Thonel\inst{2}   \and
Valérie Mezger\inst{2}  \and
Auguste Genovesio\inst{1 \scalebox{0.6}{\faEnvelopeO}} }


%
\authorrunning{Under review}
%
\institute{ENS \inst{1}, Université de Paris Cité \inst{2} \\
} 
\maketitle              
\begin{abstract}
For the past few years, deep generative models have increasingly been used in biological research for a variety of tasks. Recently, they have proven to be valuable for uncovering subtle cell phenotypic differences that are not directly discernible to the human eye. However, current methods employed to achieve this goal mainly rely on Generative Adversarial Networks (GANs). While effective, GANs encompass issues such as training instability and mode collapse, and they do not accurately map images back to the model's latent space, which is necessary to synthesize, manipulate, and thus interpret outputs based on real images. In this work, we introduce PhenDiff: a multi-class conditional method leveraging Diffusion Models (DMs) designed to identify shifts in cellular phenotypes by translating a real image from one condition to another. We qualitatively and quantitatively validate this method on cases where the phenotypic changes are visible or invisible, such as in low concentrations of drug treatments. Overall, PhenDiff represents a valuable tool for identifying cellular variations in real microscopy images. We anticipate that it could facilitate the understanding of diseases and advance drug discovery through the identification of novel biomarkers.
\keywords{generative models \and subtle phenotypes \and microscopy images.}
\end{abstract}

\renewcommand{\thefootnote}{\relax}\footnotetext{\scalebox{0.6}{\faEnvelopeO} Correspondence: auguste.genovesio@ens.psl.eu}
\renewcommand{\thefootnote}{\relax}\footnotetext{*  Equal contribution}

\section{Introduction}
The emergence of automated imaging and high-throughput platforms has made image-based cellular profiling essential for identifying phenotypic variations~\cite{cellular_profiling,cellular_profiling_2}. Traditional methods such as cell segmentation and quantification of characteristics such as intensity, shape, and texture, are commonly used to identify cellular changes in microscopy images \cite{cellprofiler}. However, these feature shifts are often challenging to interpret. This difficulty is compounded by the inherent variability among cells within the same condition, which can mask the differences between distinct conditions and complicate accurate analyses \cite{phenexp}. Detecting subtle visual differences between cells in biological images poses a significant challenge but also offers substantial potential for advancing disease understanding, discovering new biomarkers, and developing drugs and diagnostics \cite{drug_profil,cellular_profiling_3}. Recently, generative models have been explored to uncover and analyze cellular phenotypes in microscopy images \cite{phenexp,unbiased_cell,bourou}.

In \cite{bourou}, the approach to identify cellular variations was framed as an image-to-image translation task between 2 classes, similar to methodologies found in \cite{cyclegan,pixtopix}.  The core concept of this method involves training GANs to translate images between two conditions. However, typical High Content Screening (HCS) campaign test for a vast range of conditions, such as different concentrations of potential drugs. This complexity makes the method described in \cite{bourou} less practical for scenarios involving more than 2 conditions.

In Phenexplain~\cite{phenexp}, the authors proposed training a conditional StyleGAN \cite{stylegan2} to identify cellular changes by performing interpolation in the latent space. Unlike the approach in \cite{bourou}, this method accommodates training across multiple conditions. However, a significant limitation is that cellular changes are identified on \emph{synthetic} images rather than real ones, which may limit the method's applicability. A similar approach was presented in \cite{unbiased_cell}; however, instead of leveraging the latent space of GANs, the authors proposed learning a representation space using self-supervised learning techniques \cite{self_supervised_survey}.

These methods commonly employ GANs, which are known to suffer from limitations such as training instability and mode collapse \cite{mode_collapse}. Recently, Diffusion Models (DMs) \cite{ddim,ddpm,diffusion_beat_gans} have emerged as the new standard in the field of generative models, successfully addressing many of the challenges associated with GANs.

In this work we introduce \PhenDiff{}: a novel approach utilizing \emph{multi-class} conditional DMs to translate \emph{real} cell images to other conditions, allowing to spot subtle phenotypic differences triggered by a perturbation. Our code is openly available  \textbf{\href{https://github.com/WarmongeringBeaver/PhenDiff}{on GitHub}}.

\section{Methods}

\begin{figure}[htbp]
\centering
\includegraphics[width=\textwidth]{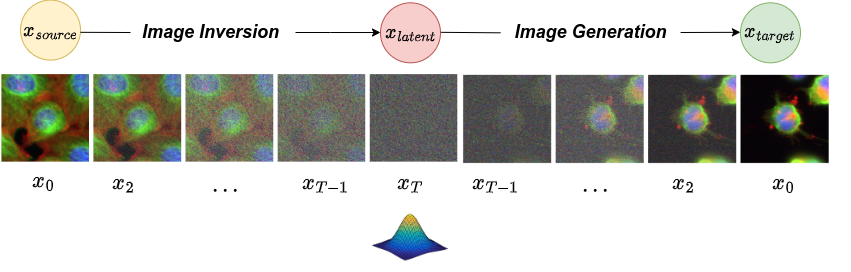}
    \caption[desc]{On this example of the BBBC021 dataset, we translated an 
    untreated image to a treated image with the highest concentration of Latruculin B, notice how the model is able to reproduce the phenotype of the target condition (lower cell count due to the toxicity, no actin cytoskeleton (red channel)) on the real image.}
\label{fig:phendiff}
\end{figure}

\PhenDiff{} is built on \textit{Denoising Diffusion Implicit Models} (DDIMs) \cite{ddim}. It comprises two stages: image inversion and image generation, as shown in Fig. \ref{fig:phendiff}. A similar approach was proposed in DDIBs \cite{ddib} where the authors proposed an image-to-image translation method that relies on two DMs trained independently on each domain. In our approach, we train a \emph{single}, \emph{conditional} DM on multiple domains simultaneously. In this section we first provide an overview of DMs and then dive in the details of our approach.

\subsection{Background}
\textit{Denoising Diffusion Probabilistic Models} (DDPMs) are one of the earliest and most successful DMs. They are latent variable models that make use of two Markovian processes: a fixed forward process that turns data into noise and a learned backward process that turns noise back into the data distribution.

\noindent Formally, given a data distribution $x_{0} \sim q(x_0) $, the forward process iteratively perturbs the data by adding  Gaussian noise to it at successive timestamps. When conditioned on $x_0$, the joint distribution of the noised images at timestamps $1,2,...,T$, denoted as $x_1,x_2,...,x_T$, can be factorized as follows:
\begin{equation}
q(x_{1:T}|x_0)= \prod_{t=1}^{T}q(x_{t}|x_{t-1})\label{eq}
\end{equation}
with the transition kernel $q(x_{t}|x_{t-1})$ given as:
\begin{equation}
q(x_t|x_{t-1})= N(x_t;\sqrt{1-\beta_{t}}x_{t-1},\beta_{t}I)\label{eq2}
\end{equation}
where $(\beta_t)$ are fixed hyperparmaters. In the backward process the noise is gradually removed by using a learnable transition kernel given by:
\begin{equation}
p_{\theta}(x_{t-1}|x_t)= N(x_{t-1}; \mu_{\theta}(x_t,t),\Sigma_{\theta}(x_t,t))\label{eq3}
\end{equation}

\noindent Similar to latent variable models, DMs can be trained using the Variational Lower Bound. In DDPMs \cite{ddpm}, the authors derived the following simplified objective function to minimize:
\begin{equation}
\mathbb{E}_{x_0\sim q(x_0), t, x_t\sim q(x_{1:t}|x_0)}{[\lVert \epsilon - {\epsilon}_{\theta}(x_t,t) \rVert_2^2]}
\end{equation}
where $\epsilon_{\theta}$ is a learnt function that predicts the noise $\epsilon \sim \textit{N}(\textbf{0},\textbf{I})$ added to $x_{t}$ by the forward process, $\epsilon_{\theta}$ is often parmeterized by a UNet \cite{Unet} network. DDPMs require many iterations at inference time to generate satisfying images. To speed up the inference, Denoising Diffusion Implicit Models (DDIMs)~\cite{ddim} can be utilized. Importantly, DDIMs also enjoy another compelling property: deterministic sampling. This allows \emph{exact} inversion, which is instrumental in our approach where we aim at observing phenotypic changes in real images.

\subsection{Deterministic Conditional Image Generation}
Sampling from a DM corresponds to gradually removing noise from noised images. As described in \cite{inver,ddim},
when using the DDIMs deterministic formulation, given $x_t$, a noised image at timestamp $t$, the denoised version of it at the  timestamp $t-1$ is given by the following formula:
\begin{equation}
x_{t-1}= \sqrt{\frac{\alpha_{t-1}}{\alpha_{t}}}x_{t} + \sqrt{\alpha_{t-1}}\gamma_t\,\epsilon_{\theta}(x_t,t,y)
\label{eq:cond_DDIM}
\end{equation}
where  $\gamma_t=\left(\sqrt{\frac{1}{\alpha_{t-1}}-1}-\sqrt{\frac{1}{\alpha_t}-1}\right)$, $\alpha_t=\prod_{i=1}^{t}(1-\beta_{i})$, and $\epsilon_{\theta}(x_t, t, y)$ is the predicted noise. We repeat this operation starting from $x_T$, which corresponds to pure Gaussian noise, to $x_0$, which is the generated image. Conditional generation is achieved by giving the class label $y$ as additional input to $\epsilon_\theta$.

\subsection{Image Inversion}\label{sec:inversion}
Image inversion is the task of finding a latent code that generates back a given real image. It plays a major role in image editing models \cite{ddib,inver,diffusionclip}. GANs inversion methods are based either on optimization or on learning an image-to-latent encoder \cite{ganinversion}. Despite recent progress, GAN inversion remains challenging due to the reduced dimensionality of the latent space in comparison to the image pixel space, as opposed to DMs. In DDIMs, a unique inverted latent code can be obtained~\cite{inver,diffusion_beat_gans}, without any additional optimization or encoding network. In the limit of small steps, The inversion formula is as follows:

\begin{equation}
x_{t+1}= \sqrt{\frac{\alpha_{t+1}}{\alpha_{t}}}x_{t} + \sqrt{\alpha_{t+1}}\Bar{\gamma}_t\epsilon_{\theta}(x_t,t,y)
\end{equation}
where $\Bar{\gamma}_t=\left(\sqrt{\frac{1}{\alpha_{t+1}}-1}-\sqrt{\frac{1}{\alpha_t}-1}\right)$.

\section{Experiments and Results}

In this section, we first present the datasets we used in our evaluation. We assess cellular variations across the conditions of these datasets both qualitatively and quantitatively. Subsequently, we compare our approach with those based on GANs. Our models were trained using 3 V100 GPUs. The network architecture is a U-Net with 3 ResNet blocks per encoder/decoder and approximatively 70M parameters.

\subsection{Datasets}

\begin{figure}[htbp]
    \centering
    \includegraphics[width=1\textwidth]{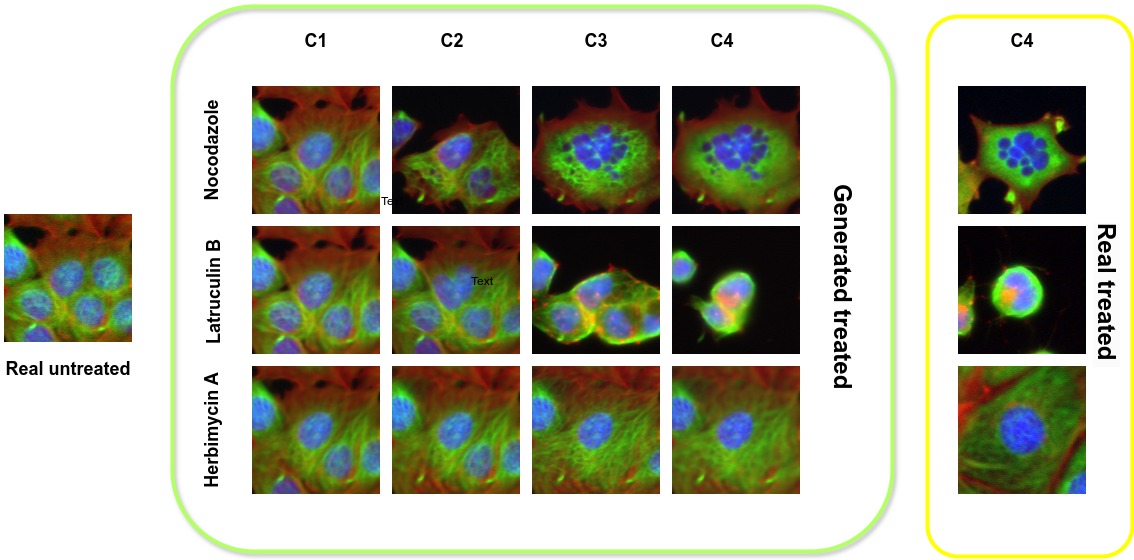}
    \caption{Translating a single real untreated image into treated counterparts for a given treatment (row) with increasing concentrations (columns, C1, C2, C3, C4). All images within the green rectangle are generated. Within the yellow rectangle are real images treated with corresponding drugs at concentration C4. The true changes induced by the highest concentration closely resemble those in the generated images at C4.}
    \label{fig:BBBC021}
\end{figure}

\noindent \textbf{BBBC021}:
The BBBC021 \cite{bbbc} is a publicly available dataset containing images obtained with  fluorescent microscopy of MCF-7 breast cancer cells treated with 113 small molecules across eight concentrations. In our research, we specifically used images of untreated cells and cells treated with 8 concentrations of of 3 compounds: Latrunculin B, Nocodazole, and Herbimycin A (25 conditions altogether). 
In Fig. \ref{fig:BBBC021}, the green, blue and red channels label for B-tubulin, DNA and F-actin respectively.

\noindent \textbf{Golgi}:
Fluorescent microscopy images of HeLa cells untreated (DMSO) and treated with Nocodazole. In Fig. \ref{subfig:golgi-first}, the green and blue channels label for B-tubulin and DNA respectively.
\subsection{Reliable synthesis of visible cell phenotypes}

\begin{figure}[ht!]
    \centering
    \begin{subfigure}[b]{0.4\linewidth}
        \centering
        \includegraphics[width=\linewidth]{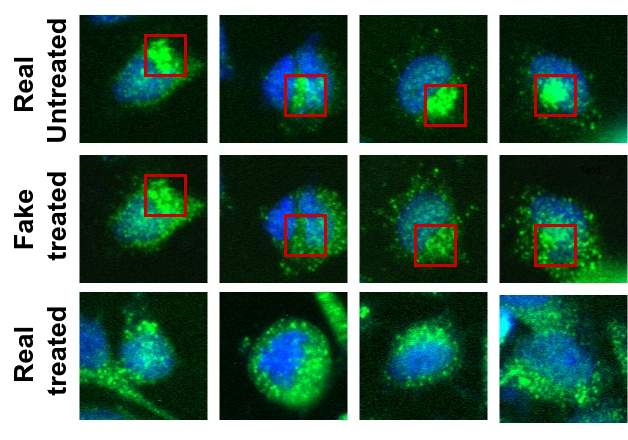}
        \caption{Translation from untreated images to treated ones.}
        \label{subfig:golgi-first}
    \end{subfigure}
    \hfill
    \begin{subfigure}[b]{0.5\linewidth}
        \centering
        \includegraphics[width=\linewidth]{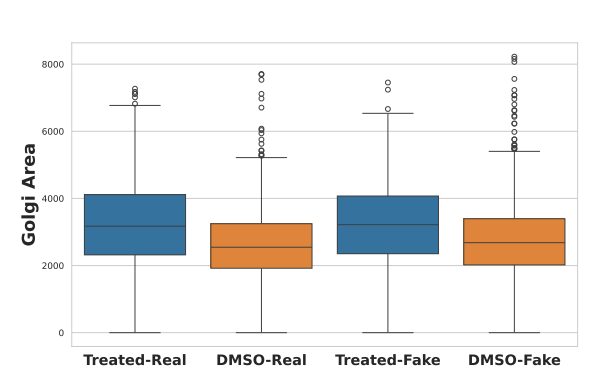}
        \caption{Boxplot of the Golgi area (in pixels).}
        \label{subfig:golgi-second}
    \end{subfigure}
    \caption{We translated real untreated images from the Golgi dataset to treated ones, we can see that Phendiff is able to replicate the effect of the treatment.
    Left: the Golgi apparatus (red box) is compact in untreated images and splitted in the treated ones (real and generated). Right: An image analysis measurement (Golgi apparatus area) performed on real and synthetic images of both conditions led to the same quantitative conclusion. A two-sided t-test yields a p-value of 1.1e-28 for real images and a p-value of 1.1e-14 for generated images. }
    \label{fig:golgi}
\end{figure}

In Fig \ref{subfig:golgi-first}  we translate real images of untreated cell to treated images with PhenDiff. We can observe changes in the true morphology of the Golgi apparatus following treatment with Nocodazole: the apparatus has fragmented into smaller stacks. PhenDiff is able to generate translated images that match this phenotype. To validate this observation quantitatively, we computed the area (measured in pixels) occupied by the the Golgi apparatus (green channel). In Fig \ref{subfig:golgi-second} we can see that there is a significant difference in the mean areas between untreated and treated cells in real images. The same difference is observed in the generated images, which indicates that our method is able to replicate the effects of this treatment and would lead to the same conclusion (see Appendix B for more examples) .

\subsection{Reliable synthesis of invisible cell phenotypes}
In the previous section, we demonstrated that the obvious phenotypic changes triggered by a treatment could be reliably reproduced. To prove that this method is also valid for detecting invisible phenotypic variations, we translated untreated real images into different classes, each corresponding to a treatment with a given concentration: [C1, C2, C3, C4]; specifically, the concentrations in $\mu M$ are the following: for Nocodazole $[0.003, 1.0, 3.0, 30.0]$, for Latruculin B $[0.003, 1.0, 3.0, 30.0]$ and for Herbimycin $[0.003, 0.3, 1.0, 10.0]$. We extracted 215 features from 1,000 translated (to all the conditions) and real images using CellProfiler \cite{cellprofiler}. For each feature, we calculated the mean values for both the real and generated images for each condition (a condition being a treatement at one concentration). We then computed the correlation between the real and synthetic mean values for that specific feature across all concentrations of a given treatment (see Appendix A). Fig. \ref{fig:CellProf}(a)-(c) display histograms of the correlation values, showing that, for the large majority of CellProfiler features, there is a strong correlation between the generated and real images. This indicates that our method can faithfully reproduce cellular variations across all concentrations including the lowest ones, for different treatments, demonstrating that synthetic images displaying invisible phenotypes can also be accurately reproduced.

\subsection{Subtle phenotypic variation can be identified on low concentrations}
 Fig. \ref{fig:BBBC021} shows that the generated images of low concentrations of distinct treatments are slightly different. As we increase the concentrations, some cells at the border of the images are systematically eliminated due to the toxicity of the treatments, although not in the same way and at the same concentrations. Moreover, Latrunculin B tends to contract the cytoplasm, whereas Nocodazole tends to extend it with increasing concentrations, something hardly visible in real images. At the highest concentration, the phenotype changes induced by the treatments closely resemble those observed in the real images.

\begin{figure}[htbp]
    \centering
    \includegraphics[width=1\textwidth]{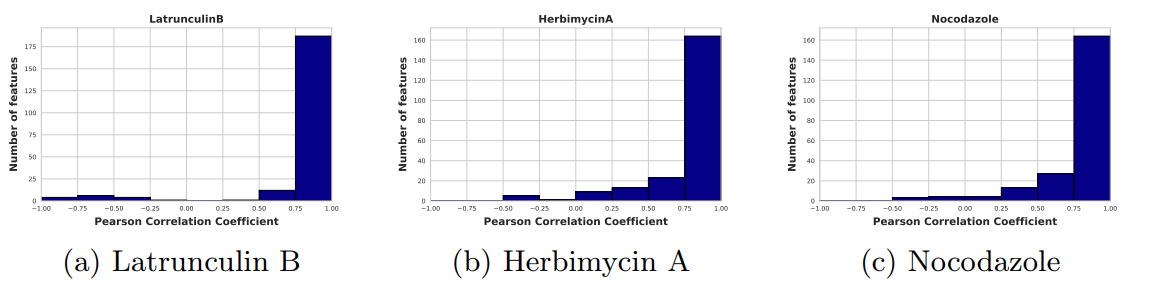}
    \caption{Distribution of Pearson correlations coefficients between the features extracted in real and translated images for each treatment. We observe that the majority of features in the real and the translated images are strongly correlated which indicates that the proposed method is able to recover the cellular variations in the microscopy images.}
    \label{fig:CellProf}
\end{figure}

\begin{table}
\caption{The FID scores for the translated images to all the concentrations of Latrunculin B treatment, the lower the score the better. }\label{tab1}
\begin{tabular}{|l|l|l|l|l|l|l|l|l|}
\hline
Method & C1 & C2 & C3 & C4 & C5 & C6 & C7 & C8 \\
\hline
StarGAN &  91.17& 86.57 & 89.39 & 86.57 & 108.53 & 114.89 & 119.47 & 111.72 \\
CycleGAN & 31.59& 30.7 & 28.4 & 34.18 & 32.85 & 21.53 & 20.38 & 24.57 \\
PhenEx/W & 25.07 & 24.6 & 22.31 & 28.93 & 40.6 & 91.86 & 131.49 & 137.95 \\
PhenEx/W+ & 36.48 & 36.62 & 31.29 & 42.80 & 58.17 & 148.32 & 203.15 & 220.63 \\
PhenDiff (ours)  &  \textbf{22.19}& \textbf{21.18} & \textbf{22.19} & \textbf{21.41} & \textbf{23.37} & \textbf{18.38} & \textbf{14.26} & \textbf{16.6} \\
\hline
\end{tabular}
\end{table}

\subsection{Evaluation of PhenDiff against existing methods}

We recall that Phenexplain~\cite{phenexp}, a method leveraging conditional StyleGAN, was introduced for identifying subtle phenotypic variations \cite{phenexp}. However, its application to real images is limited. 
To address this, we adapt it for use with real images.
This adaptation involves mapping real images into the latent space of a pre-trained StyleGAN model. We identify the vector representing the transition from the image's current class to a target class by calculating the difference between the average latent codes for each class. This allows us to move the image's latent code along this vector to produce the translated image.

Adapting Phenexplain for real images required StyleGAN inversion, a step marked by challenges. 
The inversion quality, essential for our task, varies depending on the use of the original W or extended W$^+$ latent spaces,
each offering different trade-offs between reconstruction fidelity and editability~\cite{abdal2019image2stylegan,wulff2020improving}. Acknowledging the trade-offs between using the W and W$^+$ spaces, we developed two versions of the adapted method: PhenEx/W and PhenEx/W+.

\begin{table}
\caption{Comparing the quality of the reconstructed real untreated image using the L2 loss}\label{tab2}
\begin{tabular}{|l|l|l|l|l|l|l|l|l|}
\hline
Method & StarGAN & PhenEx/W & PhenEx/W+ & PhenDiff\\
\hline
Reconstruction Loss  &  1986.80& 2001.09 & 1895.58 & \textbf{415.15}  \\

\hline
\end{tabular}
\end{table}

Additionally, considering the use of a variant of CycleGAN for the task of identifying subtle phenotypic variation in real images in \cite{bourou}, we also included CycleGAN as another baseline for comparison. However, this model is limited to translations between no more than two classes. To evaluate our method against a model capable of multi-class translations, we included StarGAN \cite{stargan}, a representative method for multi-domain image-to-image translation, into our baseline comparisons.
For the evaluations, PhenDiff was trained on all the concentrations of Latrunculin B, a process replicated for StarGAN. Due to its limitation in supporting multi-domain image-to-image translation, CycleGAN necessitated the training of eight different models, each enabling translation between the untreated class and another concentration.

In Table \ref{tab1}, we applied our method to translate 1,000 untreated images to the 8 concentrations of Latrunculin B, evaluating the generated images quality using the Fréchet Inception Distance (FID) score \cite{FID}. StarGAN, with its design focus on natural images, exhibits high FID values across all classes, suggesting its lower effectiveness for biological image translation. CycleGAN demonstrates moderate translation quality with acceptable FID scores but requires training eight different models for each treatment, making it computationally intensive.
PhenEx/W shows better performance at lower concentrations compared to PhenEx/W+, yet both struggle with accurately replicating effects at higher doses, as indicated by increasing FID scores.

The quality of the reconstructed real untreated images, as shown in Table \ref{tab2}, is crucial for our analysis. Here, PhenDiff stands out by achieving the lowest reconstruction loss, emphasizing its enhanced ability to detect cellular variations in real images (see Appendix C).
The development of PhenEx/W and PhenEx/W+ was motivated by the challenges of GAN inversion, seeking to balance reconstruction fidelity and editability. Despite this, the marginal difference in their reconstruction losses suggests a subtle balance in their capabilities. Overall, these results highlight our method's superiority in image generation quality and inversion accuracy compared to the baseline models, demonstrating its effectiveness in handling the complexities of biological image translation.


\section{Conclusion}
In this work, we introduced PhenDiff, a multi-class image-to-image translation method leveraging conditional diffusion models to identify subtle phenotypic variations in real microscopy images. Our experiments demonstrate that PhenDiff can accurately produce variations in phenotypes induced by various treatments. Additionally, compared to existing methods, particularly those based on Generative Adversarial Networks (GANs), PhenDiff exhibits superior performance in terms of image quality. Moreover, its precise image inversion capability enables the detection of these variations in real images. Overall, our findings suggest that PhenDiff can be a valuable tool in understanding the effects of certain treatments and in identifying new biomarkers.

%
%
%
\bibliographystyle{unsrt}
\bibliography{Paper-1948}
\end{document}